*Feature Article*

We develop approaches to realizing feed-forward loops and certain memory functions (associative memory) with autonomous enzymatic-cascade biochemical processes. Such systems will offer possibilities for utilizing bio-inspired information handling steps on par with the presently used digital gates. Thus, we can be guided by nature's mechanisms in our experimenting with new information and signal processing designs.

# Can bio-inspired information processing steps be realized as synthetic biochemical processes?

**Vladimir Privman**[*,1] **and Evgeny Katz**[**,2]

[1] Department of Physics, Clarkson University, Potsdam, NY 13699, USA
[2] Department of Chemistry and Biomolecular Science, Clarkson University, Potsdam, NY 13699, USA

* Corresponding author: e-mail privman@clarkson.edu, Phone: +1 315 268 3891
** e-mail ekatz@clarkson.edu, Phone: +1 315 268 4421

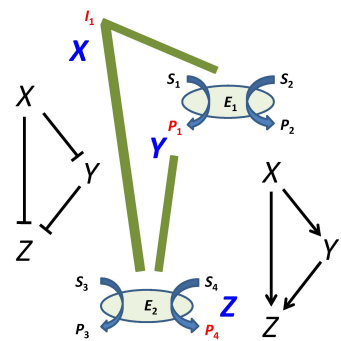

We consider possible designs and experimental realizations in synthesized rather than naturally occurring biochemical systems of a selection of basic bio-inspired information processing steps. These include feed-forward loops, which have been identified as the most common information processing motifs in many natural pathways in cellular functioning, and memory-involving processes, specifically, associative memory. Such systems should not be designed to literally mimic nature. Rather, we can be guided by nature's mechanisms for experimenting with new information/signal processing steps which are based on coupled biochemical reactions, but are vastly simpler than natural processes, and which will provide tools for the long-term goal of understanding and harnessing nature's information processing paradigm. Our biochemical processes of choice are enzymatic cascades because of their compatibility with physiological processes in vivo and with electronics (e.g., electrodes) in vitro allowing for networking and interfacing of enzyme-catalyzed processes with other chemical and biochemical reactions. In addition to designing and realizing feed-forward loops and other processes, one has to develop approaches to probe their response to external control of the time-dependence of the input(s), by measuring the resulting time-dependence of the output. The goal will be to demonstrate the expected features, for example, the delayed response and stabilizing effect of the feed-forward loops.

## 1 Introduction

Information and signal processing with biomolecules (termed "biocomputing" for brevity) stands out as an active basic-research field [1-6] in the broader context of chemical [7-11] unconventional computing approaches [12, 13]. It promises capabilities [4] to develop novel approaches to biosensing and to interfacing Si electronics for biocompatibility with living organisms. Furthermore, biocomputing can offer tools for developing information processing paradigms other than those presently widely used in analog/digital devices. One example of a recent success has been ideas [14-16] of using bio-inspired memory elements (memristors, etc.) for novel designs of Si electronic circuitry for specific applications.

This suggests that it would be of interest to devise synthetic *biochemical* systems which are simple realizations of the nature's information and signal processing functionalities, to have building blocks to experiment with in order to better understand the nature's and also enrich the analog/digital information processing paradigms.

Here we work in the framework of biocomputing biomolecular systems based on enzymatic cascades [17-22]. Biomolecular computing has been studied by many research groups, using synthetic DNA chains (oligonucleotides) [1, 5], various proteins (including enzymes) [18-25], and other bio-objects [26-28] (even whole cells) designed by nature. Use of enzymes has been motivated by their compatibility with physiological processes in vivo and with electronics (e.g., electrodes) in vitro, and selectivity and specificity, which allow networking. Furthermore, enzymes are abundant in all body functions and widely used of biomedical testing. Even small-scale networking for several-input/step information/signal processing with enzymes thus offers interesting applications [29, 30] for biosensing [31-34], e.g., for the point-of-care [35-38] rather than clinical testing, or for continuous monitoring for envi-ronmental and security/military applications [39, 40]. Enzymatic processes are also well suited for interfacing biocomputing steps with standard electronics [41-46] in electrochemical settings.

The analog/digital electronics fault-tolerant scalability paradigm is the only one that is presently fully understood. Thus, most research in biocomputing has been devoted to realizing digital logic, i.e., binary gates [21, 26, 29, 47-51] such as AND, OR, XOR, etc., and attempting to connect them in presently rather small networks [6, 18, 22, 24, 52], including those carrying out Boolean functions. Recent developments have focused on non-binary "network elements" that improve the noise handling by "filtering" approaches [53-63] useful for improving scalability of binary biocomputing gate networks.

The information processing paradigm of nature has also been very successful. We do not fully understand it, but advances have been made in systems biology to explore aspects of the functioning of the nature's information processing [64-67]. Presently, we cannot even remotely mimic the complexity of the natural processes by making "artificial life" systems starting with biomolecules/biochemistry. However, an interesting avenue of research has been to consider specific processes: memory, learning, etc., as "network elements" that could offer new functionalities to our otherwise more conventionally manufactured systems of Si-electronics. We already mentioned recent successes involving such ideas for novel electronic circuit designs that make uses of memory elements [14-16, 68, 69].

Here we take up a new challenge: We consider how to actually realize with biochemical processes certain basic bio-inspired information processing steps. Ultimately, such systems can offer tools for experimenting with information processing networks based on synthetic autonomous biochemical processes, to allow a new avenue for understanding the nature's information processing paradigm.

Our emphasis here will be on feed-forward loops. Indeed, it is the most common network motif in information processing in natural systems, and the challenge will be to carry it out with few coupled biochemical reactions, which will be an immensely simpler realization than that in nature. We will also consider certain memory processes, for which our recent preliminary work has indicated possibilities for designing biomolecular realizations [70, 71]. We will focus on associative memory for reasons described later. The present "concept article" describes the general principles of the design of such systems.

Both general and specific designs are presented and theoretically substantiated, with the bulk of the required experimental work to be carried out in the future by our group. It is hoped that the described designs will initiate a new research direction of synthetic information processing mimicking not the full scope of what the nature does, as "artificial life," but rather taking up a more limited and therefore hopefully more tractable goal of mimicking only the nature's information processing design, with biomolecular/biochemical reaction processes vastly simpler than those that evolved naturally.

## 2 Example and discussion of feed-forward

In order to make the presentation specific, let us first consider an example, which is later revisited and described in detail in Sec. 3, of a possible design of a feed-forward-loop function with a cascade of enzyme-catalyzed reactions. This section offers a general introduction, whereas the details of the actual biochemistry of this and other systems' functioning are explained in later sections. Unlike the various earlier-realized "biochemical gates," feed forward is in most cases not a binary function [72-74].

Another important difference in attempting actual biochemical realizations is that feed forward has two "signal transduction" steps, each involving the input signal, $X$, usually not being directly converted into the output. Rather, in the primary (direct) step the input acts as the activator (promoter), denoted by →, or repressor (inhibitor), denoted by ⊣, of the ongoing process(es) that generate the output signal, $Z$: See the schematic in Fig. 1. The feed-forward loop is completed by adding the secondary (indirect) step in which $X$ activates or represses the ongoing process(es) which generate another, intermediate signal, $Y$, which in turn activates or represses the process(es) of the production of $Z$.

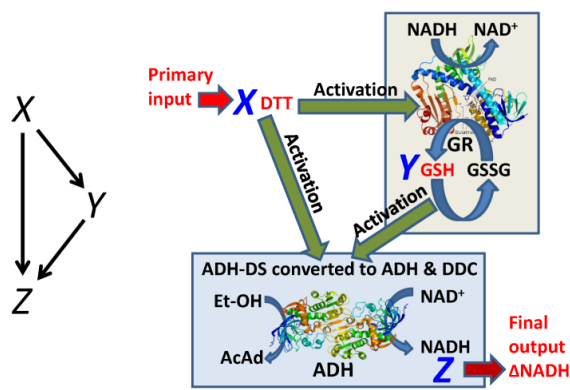

**Figure 1** Feed-forward system with *activation* in all the signal transduction steps. The schematic (left) shows the activations involved. The biochemical processes (right) are explained in detail in Sec. 3. Two enzyme-biocatalyzed processes (outlined by boxes) continuously produce chemicals which are signals $Y$ and $Z$. Chemical input $X$ activates (promotes) the production of both $Z$ and $Y$, whereas $Y$ promotes the production of $Z$. (The chemical notations and abbreviations are explained in Sec. 3.)

Here we will consider the simpler (for biochemical realizations) situation when $X$ or $Y$, rather than $X$ and $Y$ together — which is another feed-forward option — activate the output signal Z production. Then in the sim-simplest classification [74, 75] there can be 8 different loops corresponding to choosing activation or repression in each of the steps

$$X \to Z \text{ or } X \dashv Z, \quad X \to Y \text{ or } X \dashv Y, \quad Y \to Z \text{ or } Y \dashv Z. \quad (1)$$

The feed-forward loop is then called "coherent" or "incoherent" depending on whether the net effect of $X$ on the production of $Z$ in the secondary step is the same as in the primary.

The most abundant in nature [75-77] feed forward involves three activations. In Fig. 1 we show a potential realization of such a process with an enzymatic cascade, for which the notation, details of the biochemical processes, and the system's functioning are explained later, in Sec. 3, as part of a discussion of experimental realizations, including also systems involving repression.

Here we would like to address several important features expected of such systems. First, we need at least two enzymatic (or other biochemical) processes which yield signals $Y$ and $Z$, and these processes' rates at time $t$, should be affected (controlled) by the value of the input at that time, $X(t)$. In some situations activation can be made rather sharp as a function of parameters, and inhibition can also be made sharp. Thus, various responses of the feed-forward loop can be made quite steep. Feed forward can therefore on its own in some regimes approximate binary gates. Therefore, ideas have been developed [78] for multi-gate logic with DNA-structure oligonucleotide systems made of binary feed-forward functions.

However, in a general setting the feed forward's role in nature is obviously not binary. Rather, the feed-forward loop, specifically the one with three promotions shown in the schematic in Fig. 1, plays a stabilizing role in nature's networks: It delays [76, 79-81] the changes in the response, $Z(t)$, to avoid erratic swings and "waste of resources" in natural-pathway responses to environmental variations/fluctuations, specifically, those in $X(t)$. The secondary step — with the $X$ to $Y$ to $Z$ transduction — takes a fraction of the input signal and processes it in parallel to the primary transduction channel. It should be designed to act to time-delay (as compared to the direct $X$ to $Z$ transduction in the primary step) the effect of a part of the changes in the input signal, $X$, as far as its net impact on Z goes.

Therefore, in order to experimentally accomplish proper feed-forward realizations, we need to go beyond the "input at $t = 0$ to output at gate time $t_{gate} > 0$" response paradigm of the digital-gate biocomputing. We have to experimentally control the availability of $X(t)$, by adding/removing (inputting/deactivating) this compound by physical or (bio)chemical means at the externally controllable rate (which can be negative) $R_{ext}$, such that

$$\frac{dX}{dt} = R_{ext}(t) + \text{reaction terms}, \quad (2)$$

where the "reaction terms" describe the kinetics of a possible consumption of $X$ by the biochemical processes of the feed-forward loop itself. The quantification of the feed-forward effect will consist of observing how the resulting time dependence of $Z(t)$ is affected by the presence of the secondary transduction step, $X \to Y \to Z$, which can be enabled at various degrees of activity controlled by chemicals needed for that step's functioning.

One expected effect is that sharp variations in $X$ will not cause an immediate response by changes in $Z$. Rather, with the second step active at proper levels, the response will only occur when the input signal, $X$, changes (up or down) by certain threshold amounts, and furthermore, the system might have its own time scales for response rather than be driven by the input's variation.

The "stabilizing/resource-conserving" effects expected of feed-forward functions have never been realized in simple enzymatic biochemical systems. We hope to highlight the challenges involved in such realizations in this work, as well as consider possible approaches to accomplish not only feed-forward but also other bio-inspired processes, starting with those involving memory. For these the experimental realizations will have to be designed guided by the needs of quantifying the anticipated characteristics of their response. We also comment on the chemical-kinetics modeling of the expected systems.

## 3 Enzymatic feed-forward loops

In this section we present the potential enzymatic-cascade-based feed-forward designs. Let us first describe in detail the system shown in Fig. 1, outline its design, and report preliminary experimental observations. The system involves only activations in all its signal-transduction steps corresponding to the options in Eq. 1. The cascade includes the functioning of two enzymes as biocatalysts: Glutathione reductase (E.C. 1.8.1.7), abbreviated GR, which biocatalytically converts glutathione from its oxidized form, GSSG, to reduced form, GSH. The latter, GSH, acts as our intermediate signal, $Y$, in the feed-forward functioning. Concomitantly, β-nicotinamide adenine dinucleotide is converted [82] from its reduced form, NADH, to the oxidized form, $NAD^+$. Alcohol dehydrogenase (E.C. 1.1.1.1), ADH, biocatalytically oxidizes ethanol (Et-OH) to yield acetaldehyde (AcAd), while β-nicotinamide adenine dinucleotide is converted [83] from its oxidized form, $NAD^+$, to the reduced state NADH. These two processes can yield the net increase in the amount of NADH that can be measured optically by changes in absorption and that will be designated as our output signal, $Z$. Thus, as expected for feed forward, signals $Z$ and $Y$ are generated continuously once the reactions are started. In fact, the net rate of production of NADH in this system must be kept in check, to avoid rapid build-up of the signal Z. This can be done [84] by initially largely inhibiting the activity of enzyme ADH by adding disulfiram, DS, which forms an ADH-DS complex that has low biocatalytic activity.

The input signal, *X*, can then be dithiothreitol, DTT, that, when added to the system has the following effects on the process rates:

(i) Its promotion of GR results in a substantial increase [85] of the GR enzyme activity, thus increasing the rate of the signal *Y* generation. This corresponds to *X*→*Y* in Eq. 1.

(ii) In addition, DTT chemically converts DS from its original disulfide form to the thiol form, diethyldithiocarbamate, DDC, shifting the kinetics to result in the breakup of the inhibited-enzyme complex ADH-DS and restoring ADH to high activity, thus increasing the rate of the signal *Z* generation: *X*→*Z* in Eq. 1.

(iii) Importantly, GSH (signal *Y*), which is the product of the reaction biocatalyzed by GR, also chemically removes DS from the ADH-DS inhibited complex, acting to increase the rate of the signal *Z* generation. Therefore, the step *Y*→*Z* in Eq. 1, is built into the biochemical system's functioning.

The excess concentration of NADH,

$$\Delta \text{NADH}(t) = \text{NADH}(t) - \text{NADH}(0), \qquad (3)$$

which is our signal *Z*(*t*), is generated at a rate increased by the direct effect of the input signal, *X*. And it is also increased through the indirect step of *X* accelerating the signal *Y* production, while the latter in turn contributes to increasing the output of *Z*.

The full realization and characterization of the proposed enzymatic cascade will require addressing several challenges, even though some of the processes have already been studied in the literature. The latter include the process of removing the DS inhibitor from ADH-DS complex by DTT (signal *X*). Indeed, DTT is an established reactant for converting disulfide chemical species to their thiol derivatives [86]. The inhibition of ADH by the disulfide form of DS has also been studied, including the property that its thiol derivative, DDC, is removed from the enzyme complex and does not inhibit ADH [84,87]. Therefore, the primary step, *X*→*Z*, should be realizable in a controllable fashion for experiments to probe the time dependence properties of the process.

Regarding the secondary step, for *X*→*Y* we already mentioned that DTT is known to promote [85] the activity of GR. As a test of feasibility of realizing the *Y*→*Z* step, preliminary experiments were performed, see Fig. 2, to demonstrate the effect of GSH (signal *Y*) on the rate of NADH production (contributing to signal *Z*). First, ADH (0.63 units mL$^{-1}$) was prepared in the inhibited form by incubation with the added optimized concentration of disulfiram (0.8 mM). This disulfiram concentration caused a significant inhibition of ADH, by forming the ADH-DS complex. ADH could then be reactivated by the thiol-disulfide exchange with the reduced glutathione produced *in situ* by the GR reaction. Lower disulfiram concentrations did not result in a substantial inhibition of ADH, while higher concentrations did not allow the enzyme reactivation. The inhibited ADH was tested for the production of NADH in the presence of NAD$^+$ (1.0 mM) and ethanol (1.7 M). Despite its inhibition, the enzyme ADH demonstrated some activity producing NADH, Fig. 2A, curve a. Addition of NADH (50 μM) initially, to the solution in the absence of the disulfide reducing system (GR and GSSG) did not noticeably affect the rate of the biocatalytic production of the excess NADH, Fig. 2A, curve b. The same experiment performed in the presence of GR (2 units mL$^{-1}$), GSSG (3 μM) and NADH (the same added amount, 50 μM) resulted in an enhanced production of NADH, witnessing the inhibitor removal and ADH reactivation, Fig. 2A, curve c. The rate of the NADH production was increased approximately two-fold, Fig. 2B, demonstrating the feasibility of the *Y*→*Z* step.

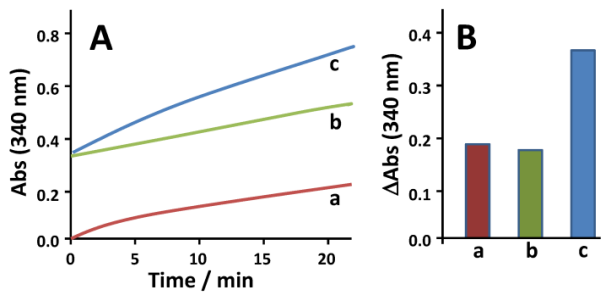

**Figure 2** Experimental probe of the feasibility of processes needed for the *Y*→*Z* step of the feed-forward loop shown in Fig. 1. Panel A: Absorbance at $\lambda_{max}$ = 340 nm, as a function of time, measuring the amount of NADH: (a) produced only with the inhibited ADH-DS enzyme complex; (b) the same with some initially added NADH, but in the absence of the other chemicals (GR and GSSG) needed for the signal *Y* production; and (c) in the presence of GR and GSSG (and the same quantity of the initially added NADH). Panel B: ΔAbs = Abs(*t*) – Abs(0), at *t* = 20 min.

Regarding the interconnectivity of the system just described (Fig. 1) with other biocatalytic processes for attempting networking, we note that the output, NADH, is a substrate (one of the input chemicals) common for many enzymes, mostly for dehydrogenases. The input, dithiothreitol is less common in purely enzymatic reactions, but it can be produced by enzyme-catalyzed processes [88]. For example, enzyme PDI (protein disulfide-isomerase, E.C. 5.3.4.1) can produce DTT in its reduced form that is needed to initiate the processes shown in Fig. 1. We also mention that the cascade in Fig. 1 can be realized with the initial NADH replaced by a different chemical, NADPH (which is converted to NADP$^+$, nicotinamide adenine dinucleotide phosphate, by the enzyme GR). Then the whole amount of the NADH produced by ADH (or ADH-DS) at time *t* > 0 can be identified as the output signal *Z*(*t*).

Generally, for nearly any enzyme there are compounds that can promote or inhibit that enzyme's activity. The latter effect can be used to devise systems that involve feed forward with some or all of the processes in Eq. 1 corresponding to repression instead of activation. As an illustration, Fig. 3 shows such a design of a system with all three processes being repressions. Two enzymatic reactions are ongoing: Glutathione oxidase, GlutOx (E.C. 1.8.3.3), biocatalytically converts glutathione from its reduced form (GSH) to the oxidized form (GSSG), while concomitantly oxygen is converted to hydrogen peroxide [88] that is taken as signal $Y$. Another enzyme, pyruvate kinase, PK (E.C. 2.7.1.40), is converting adenosine triphosphate, ATP, to adenosine diphosphate, ADP, with the concomitant conversion of pyruvate, Pyr, to phosphoenolpyruvate, PEP. The output can be defined as the produced amount of ADP. However, to actually measure it, the ATP consumption (the decrease in its concentration) should be used to yield the output signal $Z$, since it can be measured by the standard optical assay for ATP: light emission generated by luciferin-luciferase system in the presence of ATP [89]. Note that ADP/ATP are produced/consumed stoichiometrically. Hydrogen peroxide, $H_2O_2$, produced in the first biocatalytic reaction is known as an inhibitor for PK [88], and therefore the first biocatalytic process is repressing the second one; this realizes $Y\dashv Z$. In addition, cysteine, defined as signal $X$, when added to the system as the primary input will repress both biocatalytic processes. Cysteine is a known inhibitor [88] of both GlutOx, yielding $X\dashv Y$, and PK, which corresponds to $X\dashv Z$.

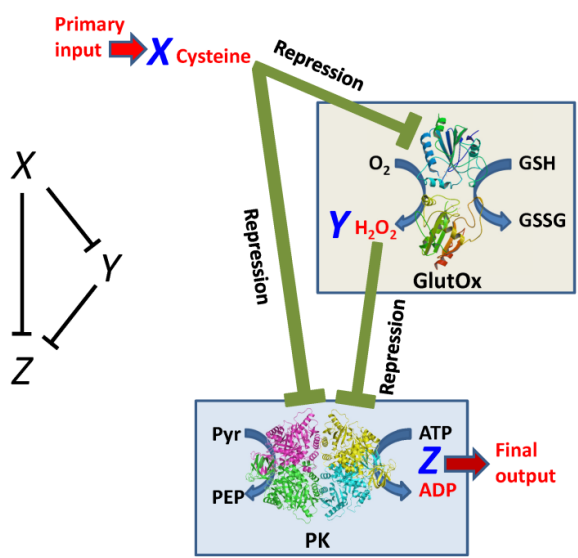

**Figure 3** Feed-forward system with repression of all the signal processing steps. Abbreviations and system functioning are explained in the text.

Regarding its interconnectivity with other biocatalytic processes, the designed system (Fig. 3) offers a substantial flexibility. Its output, ADP (or ATP) is utilized by more than 1700 ATP/ADP-dependent enzymes catalogued in the standard database [88]. The input, cysteine, is also compatible with enzymatic processes and can be produced, for example, by cysteine reductase (E.C. 1.8.1.6) [88].

For the two considered feed-forward systems (Figs. 1, 3), with all activation or repression, as well as for other possible systems with different →/⊣ combinations, modeling/design and careful selection of parameters are required for proper functioning to achieve compatible reaction conditions. Their actual experimental realization, kinetic modeling, and then the first attempts at networking with other biocatalytic steps are not straightforward and will require a substantial research effort. This expectation is based on what was learned in earlier work with digital biocatalytic gates. Indeed, to our knowledge feed-forward loops, while being extensively modeled in the literature, have never been actually experimentally realized as autonomously-functioning synthetic rather than natural biomolecular systems. Importantly for the experimental work, all the enzymes, and their substrates (input chemicals) needed for the experimental realization of the above formulated processes are commercially available.

**4 Process design and kinetic modeling**
In modeling of feed-forward loops one can set up [66] coupled phenomenological rate equations describing signal variations in the ongoing process steps. This approach can yield the expected features, including the delayed response of the output to the input's variations/fluctuations, and other properties [74]. As explained shortly, rate equations arise naturally in our systems because of the nature of the biochemical processes involved. However, they will be more complicated and contain different terms than those considered in purely phenomenological formulations. The schematic in Fig. 4 outlines one of the possible structures generic to the proposed processes considered in Sec. 3, with random selection of notations for chemicals (some not shown, cf. Figs. 1, 3) and identifications of the signals.

Our systems involve the catalytic functions of enzymes, e.g., $E_1$ (Fig. 4). Most enzymes have several functional pathways, but for our purposes it will generally suffice [18] to use the standard Michaelis-Menten type model which focuses on the dominant mechanism, described by the following process sequence. The enzyme first binds a chemical called a substrate, say, $S_1$ to form a complex, $C$. This complex can either on its own or by binding another substrate, $S_2$ — this option is common for our system, produce the product(s) of the biocatalytic reaction, here $P_{1,2}$ (see Fig. 4), restoring the enzyme to its original form. In the chemical reaction notation, we have

$$S_1 + E_1 \underset{k_{-1}}{\overset{k_1}{\rightleftarrows}} C, \quad S_2 + C \overset{k_2}{\rightarrow} E_1 + P_1 + P_2. \qquad (4)$$

This is of course just one of the possible reactant labelings (cf. Fig. 4) as far as their role in the cascade goes. Here the arrows with rate constants above/below represent chemical processes rather than activation. The second step can usually be assumed irreversible, but the first one requires two rate constants. These process parameters, here $k_{\pm1}$, $k_2$, are generally not known individually and have to be fitted from experiments. Activation/repression can involve several mechanisms, one of which can be a complex formation, for example,

$$I_1 + E_1 \underset{r_{-1}}{\overset{r_1}{\rightleftarrows}} \overline{E_1} + W. \tag{5}$$

Here the "complex" is the modified enzyme $\overline{E_1}$ with a different activity (with larger or smaller rate constants, $\overline{k_{\pm1}}$, $\overline{k_2}$ in processes similar to those in Eq. 4), and it can be restored to the original form, $E_1$, by reacting with some other chemical, here denoted $W$.

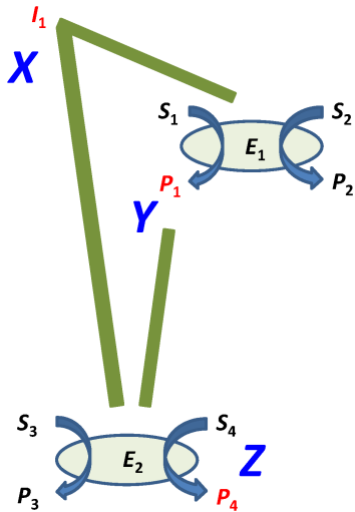

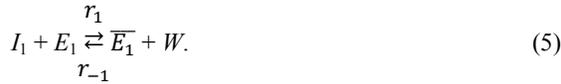

**Figure 4** Illustration of the described feed-forward designs in terms of the constituent enzymatic processes. Activations or repressions (promotions or inhibitions) are shown be green lines, cf. Figs. 1, 3.

If, for instance, $I_1$ is our input, $X$, then the added "reaction terms" in Eq. 2 enter via such chemical processes,

$$\frac{dI_1}{dt} = R_{\text{ext}}(t) - r_1 I_1(t) E_1(t) + r_{-1} W(t) \overline{E_1}(t), \tag{6}$$

whereas the time-dependence of the entering quantities is in turn set by their own rate equations, for example,

$$\frac{dE_1}{dt} = -r_1 I_1 E_1 + r_{-1} W \overline{E_1} - k_1 S_1 E_1 + k_{-1} C, \tag{7}$$

etc. Note that in the next stage, when writing the rate equation for $S_1$, for instance, terms resulting from its reaction with both the original and modified enzymes will enter, with their respective rates,

$$\frac{dS_1}{dt} = -k_1 S_1 E_1 + k_{-1} C - \overline{k_1} S_1 \overline{E_1} + \overline{k_{-1}} \bar{C}, \tag{8}$$

where the notation (such as for $\bar{C}$) is self-explanatory.

Even within a relatively simple chemical kinetics description outlined here, enzymatic cascades thus lead to systems of numerous coupled chemical rate equations, with parameters which depend on the physical and chemical conditions of the experiment, and which are documented only to a very limited extent (typically, at most a single parameter, calculated in our notation from the quantities $k_{\pm1}$ and $S_2(0) k_2$, call the Michaelis-Menten constant, is uniformly tabulated).

The described approach should enable modeling to select adjustable quantities (concentrations of those chemicals that are not designated as input/output signals) as needed to achieve expected feed-forward responses to various protocols $R_{\text{ext}}(t)$ of controlling the input signal availability, thus guiding the experimental work to achieve proper functioning of the synthetic systems, standalone and ultimately "wired" (connected via chemical process steps) with other enzymatic processes to attempt simple networking.

The main difference of the considered designs when compared to the much simpler earlier-studied binary-gate systems is the need for activation/repression as separate ongoing processes, i.e., in that enzymes are not just biocatalysts of fixed activity, but their activity changes ($E_1 \leftrightarrow \overline{E_1}$, etc.). As mentioned earlier, the need to have one of the products of the first enzymatic process activate/repress the second process is also a significant experimental design challenge.

It is therefore important to consider simpler candidate systems for feed-forward realizations specifically with all three activations, because this could be attempted by having the input signals $X$, $Y$ selected as substrates (rather than promoters) for enzymes. However, our group's preliminary modeling results (unpublished) with typical experimental as well as with theoretically optimized parameter values, indicate that the expected feed-forward-loop features are difficult to obtain in such systems and are not well pronounced. Therefore, working with the more complicated systems that have reaction structure of the type shown in Fig. 4 is warranted even for the all-promotions feed-forward case. These systems actually imitate the nature's design: Instead of $X$ being the actual direct input for the production of $Y$ and $Z$, it activates (or, if needed, represses) the already ongoing processes that produce $Y$ and $Z$, and similarly for the mechanism by which $Y$ affects the production of $Z$.

## 5 Memory Systems

Basic "learning" and "memory" steps can also be studied in the framework of the proposed approach, as outlined in this section. One example is associative memory, i.e., one signal triggering the response to another after some "training" time, which has been known starting from the Pavlov's dog experiments [90]. Various cumulative-memory "devices," with response proportional to the integrated (stored) input signal over some time interval, have recently been actively studied, including "memristors," "memcapacitors," etc. [14-16, 91-96], and the former were argued to be observable at the cellular-organism response level [92]. Preliminary solid-state experimental realizations with oxide-nanostructure-material device structures and also electrochemical setups were reported for memristors [95-99]. There is no doubt that nature uses various types of memory-involving processes at all scales, but we simply do not yet fully understand their role and the degree of their abundance at the molecular signal-processing network-structure levels. As mentioned earlier, memory device concepts have recently been found [14-16] useful in novel designs for electronic circuitry.

Our goal here is initiate consideration of the level of "complexity," cf. Fig. 4 for feed forward, required for the minimal cascade structures that are needed for realizing memory processes, and the degree of interfacing with other physical or chemical transduction steps achievable. As already mentioned, the basic-science interest in such designs is that they, including feed forward, will allow us to begin building a "toolbox" of network elements for experimenting with non-binary, bio-inspired designs for complex bio-inspired information- and signal-processing tasks.

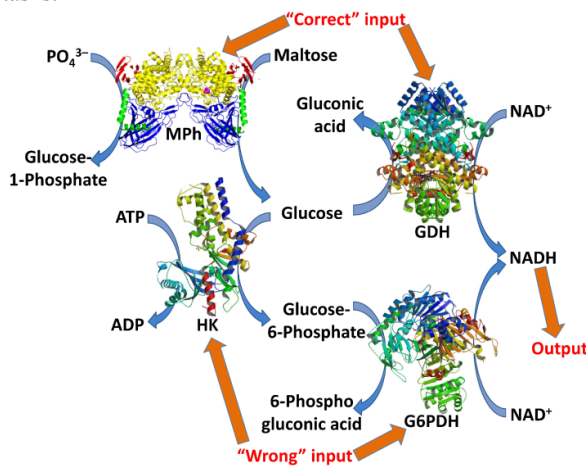

**Figure 5** Experimentally realized [70] associative memory system demonstrating learning/unlearning properties (see text). Here the "correct"/"wrong" input signals were defined as the presence/absence of various enzymes, thus limiting the possibilities for this system's autonomous integration with other biocatalytic processes for networking. (Detailed explanation of the biocatalytic reactions involved is given in Ref. [70].)

Associative memory systems were recently reported in our preliminary studies involving enzymatic processes [70,71]. A biocatalytic cascade with two parallel enzyme-catalyzed reaction branches, where one of the pathways performs the memory-needed-for-training function, was utilized to mimic the associative memory system with learning/unlearning properties, as shown in Fig. 5. However, the realized system had an important drawback: The input signals were defined as the presence/absence of various enzymes. This significantly complicates any straightforward integration into more complex networks. The system also did not allow time-dependent control of the inputs, as commented on shortly.

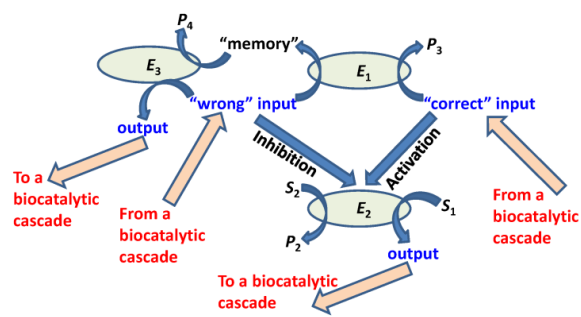

**Figure 6** Enzyme-based associative memory design demonstrating learning/unlearning properties. One of the intermediate products performs the memory function. The "correct"/"wrong" input signals are defined as the presence/absence of substrates/activators/inhibitors in the biocatalytic reactions, allowing connection of this process to other biocatalytic cascades. Here $S_{1,2}$ and $P_{1,2,3}$ denote those substrates and products in the reactions biocatalyzed by the enzymes $E_{1,2,3}$, that are not the two inputs or output signal. The "correct" input, "wrong" input, and the same output produced by both $E_2$ and $E_3$, are low molecular weight species which can be produced/consumed as products/substrates of reactions in other cascades.

Generally, for easy interconnectivity the use of low molecular weight species — those that constitute typical substrates and products of enzyme-catalyzed processes — as chemical input(s) and output signals is desirable. Indeed, our considering associative memory, rather than other bio-inspired memory "units," on par with feed forward here is based on the fact that it represents the next level of complexity: from single input (of feed forward) to two inputs, in controlling chemical concentrations the time dependence of which should be varied according to predefined protocols and the resulting response of the output studied for its time-dependence. Of course the time-dependences of the input(s) and output are quite different for the two systems, feed forward vs. associative memory, but the experimental-setup challenge in the biochemical context is similar: It entails developing capabilities for both positive and negative input-rate control, e.g., the term $R_{ext}(t)$ in Eqs. 2, 6. To enable the latter, negative rate, we need the capability to deactivate part of the input(s) by (bio)chemical processes, as further discussed below.

Figure 6 shows a schematic design, at the level of the cascade structure similar to Fig. 4, of a basic associative-memory step, realization of which can be attempted in experiments with several choices of enzymes and other chemical compounds. This system satisfies the condition of interconnectivity with other signal processing biocatalytic cascades via low-molecular-weight species as input/output signals.

Here the "correct" input will activate an enzyme ($E_2$) which converts a substrate ($S_1$) to the final output. The "wrong" input performs the opposite operation. It inhibits this enzyme resulting in no formation of the output. However, the simultaneous application of the "correct" and "wrong" inputs activates another enzyme ($E_1$), for which the "correct" and "wrong" inputs serve as substrates, resulting in the formation of an intermediate product which serves as "memory". This enables "training" such that later application of the "wrong" input alone will activate the last enzyme ($E_3$), which in the presence of the "memory" species produces the same final output as $E_2$. It should be noted that $E_3$ produces the output species only when both the "wrong" input and the "memory" species are present. The "wrong" input applied without the "memory" species earlier formed, will not result in the output formation. Similar to Ref. [70], here the associative memory is not self-reinforcing, because the rate of production of the output will decrease to zero as the "memory" compound is used up, unless the "correct" input is supplied again to "reinforce" the training.

Generally, "memory" in the considered and other recently designed [71,95] biomolecular and hybrid electrochemical systems (the latter involving interfacing with active-response electrodes) is realized by accumulation of an intermediate-product chemical compound in the cascade of processes. Direct parallels with the linear electronic circuit elements are not obvious, because chemical processes have rather different properties. In fact, ironically, even linear electronic elements that do not involve memory, specifically, a resistor, have no direct analogies in purely chemical processes. However, chemical memory has its own characteristic features expected to be useful in network designs.

As with the case of feed forward, the process realizations here are not unique, with Fig. 6 offering just one of the possible designs. Experiments will require optimization by process modeling to make sure that the realized systems correspond to the expected "training" properties as far as time dependence of the output is driven by various protocols of supplying the inputs, similar to the considerations staring with Eq. 2, Sections 3 and 4, but now with two controlled inputs instead of one.

As mentioned earlier, to accomplish this we will need to control the time dependence of the inputs, which specifically requires the capability of having both positive and negative external supply/removal rates, here for both the "correct" and "wrong" inputs. While the time-dependence protocols of interest for the two inputs here are obviously not the same as for the single input control rate, $R_{ext}(t)$, for feed forward, the experimental challenges are similar because this implies that our enzymatic cascades cannot be studied as standalone biochemical processes. They will have to be at least minimally "networked" with physical and (bio)chemical processes that add/remove (activate/deactivate) the input(s).

## 6 Conclusion

We outlined the conceptual setup for cascades of enzyme-catalyzed biochemical reactions that realize the feed-forward response or associative memory. The new aspects of the devised systems, specifically for feed forward involve the aggressive use of chemicals which are not enzymes' substrates but rather are compounds (called ligands) typically binding to other than substrate-reaction sites of the enzyme molecules and causing promotion (activation) or slowdown (repression) of the enzymatic activity. For associative memory, we focus on realizations with low-molecular-weight (rather than protein) inputs, for enabling time-dependent control.

Our presentation highlights the experimental challenges and the required modeling involved in designing, realizing and characterizing the basic steps, with feed forward as a process with a single controlled input, and associative memory with its two controlled inputs. However, at least some degree of modeling and experimentation with networking cannot be avoided: In order to probe the full range of the control by the input(s) variation, the external input rate(s) must be varied for positive values (which can be accomplished by physically adding the input reactant, as well as chemically), but also for negative values which in most cases will require an additional chemical or biochemical process for deactivation.

**Acknowledgements** Funding of our research by the NSF (grants CBET-1066397 and CBET-1403208) is gratefully acknowledged.